\documentclass{ws-procs9x6}

\begin{document}



\def\lsim{\mathrel{\raise3pt\hbox to 8pt{\raise -6pt\hbox{$\sim$}\hss {$<$}}}} 
\newcommand{\be}{\begin{equation}}
\newcommand{\ee}{\end{equation}}
\newcommand{\bea}{\begin{eqnarray}}
\newcommand{\eea}{\end{eqnarray}}
\newcommand{\da}{\dagger}
\newcommand{\dg}[1]{\mbox{${#1}^{\dagger}$}}
\newcommand{\hlf}{\mbox{$1\over2$}}
\newcommand{\lfrac}[2]{\mbox{${#1}\over{#2}$}}
\newcommand{\scsz}[1]{\mbox{\scriptsize ${#1}$}}
\newcommand{\tsz}[1]{\mbox{\tiny ${#1}$}}
\newcommand{\doref}{\bf (*** REF.??? ***)}



\title{Seeking a solution of the Pioneer Anomaly}

\author{MICHAEL MARTIN NIETO}

\address{Theoretical Division (MS-B285),\\
Los Alamos National Laboratory,\\
Los Alamos, New Mexico 87545, USA\\ 
E-mail: mmn@lanl.gov}

\author{JOHN D. ANDERSON}

\address{121 S. Wilson Ave.,\\
Pasadena, CA 9110, U.S.A.  \\
E-mail: jdandy@earthlink.net}  

\maketitle

\abstracts{
The 1972 and 1973 launched Pioneer 10 and 11 were the first missions to explore the outer solar system.  They achieved stunning breakthroughs in deep-space exploration.  But around 1980 an unmodeled force of $\sim 8 \times 10^{-8}$ cm/s$^2$, directed approximately towards the Sun,  appeared in the tracking data.  It later was unambiguously verified as not being an artifact. The origin remains unknown (although radiant heat remains a likely cause).  Increasing effort has gone into understanding this anomaly.  We review the situation and describe programs to resolve the  issue.}



\section{The Pioneer missions and their navigation}
\label{intro}
 
The first missions to fly to deep space were the Pioneers.  By using flybvs they were able to obtain heliocentric velocities that were unfeasible at the time with only chemical fuels.  
Pioneer 10 was launched on 2 March 1972 local time, aboard an Atlas/Centaur/TE364-4 launch vehicle 
It was the first craft launched into deep space, the first to reach an outer giant planet, Jupiter \cite{wolverton}, and   
the first spacecraft to leave the ``solar system." 

While in its Earth-Jupiter cruise, Pioneer 10 was still bound to the solar system.   With the Jupiter flyby, Pioneer 10 reached escape velocity from the solar system.
Pioneer 10 has an asymptotic escape velocity from the Sun of 11.322 km/s (2.388 AU/yr). [
An Astronomical Unit (AU) is the mean Sun-Earth distance, about 150,000,000 km.]    

Pioneer 11 followed soon after Pioneer10, with a launch on 6 April 1973.  It, too, cruised  to Jupiter on an approximate heliocentric ellipse.  On 2 Dec. 1974, when Pioneer 11 reached Jupiter, it underwent a Jupiter gravity assist that sent it back inside the solar system to catch up with Saturn on the far side.   It was then still on an ellipse, but a more energetic one.  
Pioneer 11 reached Saturn on 1 Sept. 1979.  The trajectory took the craft under the ring plane on approach.  After passing through the plane -- again without catastrophic consequences -- Pioneer 11  embarked on an escape hyperbolic trajectory with an asymptotic escape velocity from the Sun of 10.450 km/s (2.204 AU/yr),  
in roughly the opposite direction as Pioneer 10.

The Pioneer navigation was carried out at the Jet Propulsion Laboratory.  It was ground-breaking in its advances -- no craft had delved so far out into the solar system.  
NASA's Deep Space Network (DSN) was used to transmit and obtain the raw radiometric data.  An S-band signal ($\sim$2.11 Ghz) was sent up via a DSN antenna located either in California, in Spain or in Australia. The signal was transponded back with a (240/221) frequency ratio ($\sim$2.29 Ghz), 
and received back (at another station if, during the radio round trip, the original station had rotated out of view).  There the signal was compared with 240/221 times the recorded transmitted frequency and any Doppler frequency shift was measured directly by cycle count compared to an atomic clock. 
This produced a data record of Doppler frequency shift as a function of time, and from this a trajectory was calculated.   

However, to obtain the spacecraft velocity as a function of time from this Doppler shift is not easy.  The codes must include gravitational and time effects of general relativity, the effects of the Sun, planets and their large moons.  The positions of the receiving stations and the effects of the environment, atmosphere, solar plasma are included.  

Given the above tools, precise navigation was possible because, the Pioneers were spin-stabilized. With spin-stabilization the craft are rotated at a rate of  $\sim$(4-7) rpm about the principal moment-of-inertia axis.  Thus, the craft is a gyroscope and attitude maneuvers are needed only when the motions of the Earth and the craft move the Earth from the antenna's line-of-sight.  Thus, especially in the later years, only a few orientation maneuvers were needed every year to keep the antenna pointed towards the Earth, and these could be easily modeled.   

Even so, there remained one relatively large effect on this scale that had to be modeled: the solar radiation pressure of the Sun.  It produced an acceleration of  $\sim 20 \times 10^{-8}$ cm/s$^2$ on the Pioneer craft at the distance of Saturn (9.38 AU from the Sun at encounter).    Therefore,  any ``unmodeled force" on the craft could not be seen very well below this level at Jupiter.  However, beyond Jupiter it became possible.


\section{The Anomaly is Observed}
\label{discovery}

In 1969 one of us (JDA) had become PI of the radioscience celestial mechanics experiment for the Pioneers.   Eventually the Pioneer Doppler data going back to 1976 for Pioneer 11 and 1981 for Pioneer 10 (but also including the Jupiter flyby) was archived at the  National Space Science Data Center (NSSDC), something that later was to prove extremely helpful.  
Part of the celestial mechanics effort, working together with the navigation team, was to model the trajectory of the spacecraft very precisely and to determine if there were any unmodeled effects.   

\begin{figure}[h!] 
    \noindent
    \begin{center}  
\includegraphics[width=3.75in]{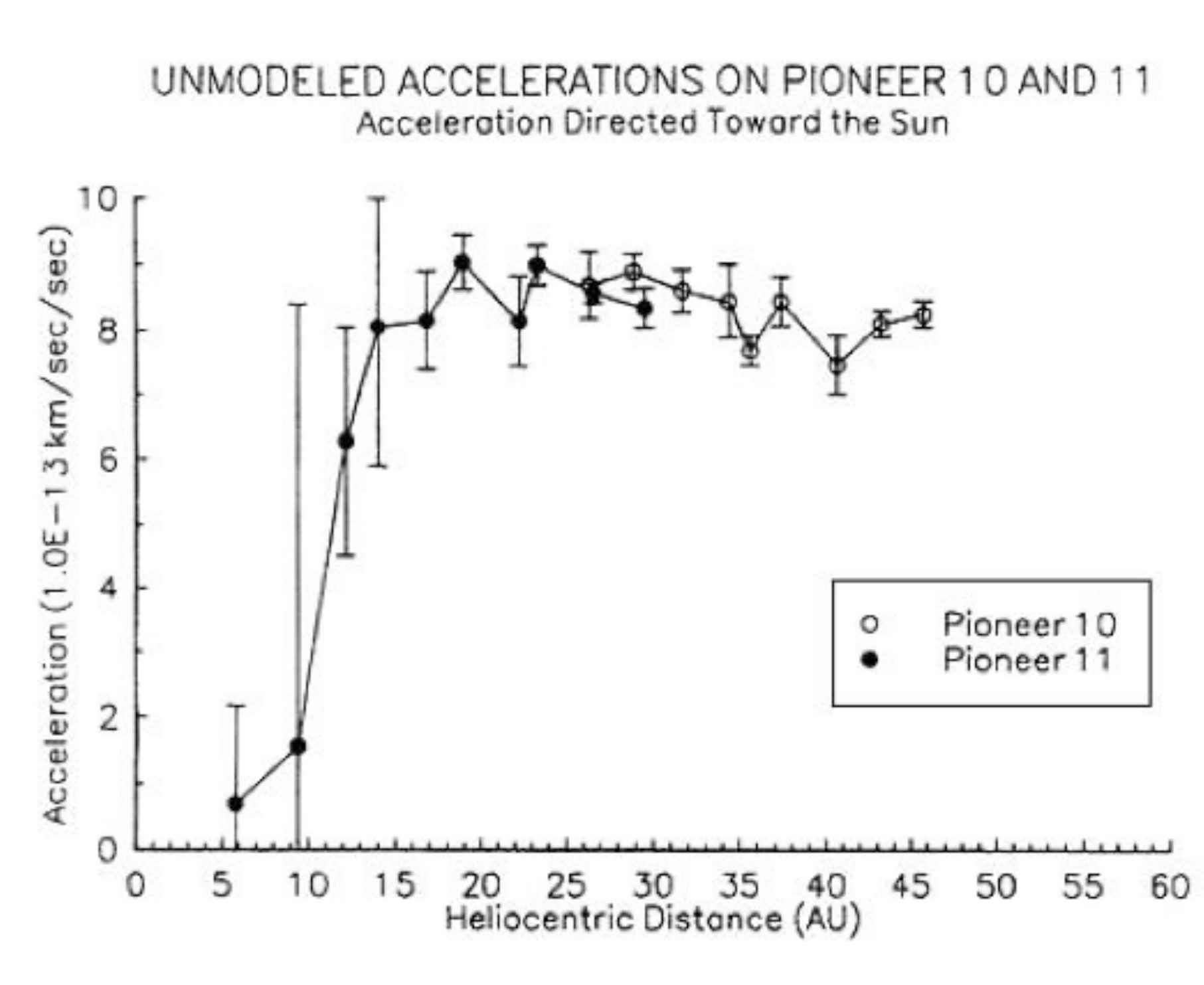}
\caption{A JPL Orbital Data Program (ODP) plot of the early unmodeled accelerations of Pioneer 10 and Pioneer 11, from about 1981 to 1989 and 1977 to 1989, respectively.  
\label{correlate} }
\end{center}
\end{figure}

After 1976 small time-samples (approximately 6-month to 1-year averages) of the data were periodically analyzed.  
These data points were obtained individually by a number of very-qualified investigators.
At first nothing significant was found.    
But when the analysis was done around Pioneer 11 's Saturn flyby, things dramatically changed.  (See the first three data points in Fig. 
\ref{correlate}.)  So people kept following Pioneer 11.  They also started looking more closely at the incoming Pioneer 10 data.  


By 1987 it was clear an anomalous acceleration $\sim 8 \times 10^{-8}$ cm/s$^2$ appeared to be acting, approximately towards the Sun    Although of concern, the effect was small and did not affect the necessary precision of the navigation.  By 1992 it was clear a more detailed look would be useful.

In 1994 MMN contacted Anderson to ask how well we understood gravity far out in the solar system.  JDA responded that, ``The biggest systematic in the (Pioneer) acceleration residuals is a bias of $8 \times 10^{-13}$ km/s$^2$ directed toward the Sun."
The result was an announcement in a 1994 proceedings.
[{\it More information and many references can be found in \cite{cphys}.}]

The strong reaction was that the anomaly might be due to JPL's ODP program, so an independent code had to test it.  Anderson put together a team that included two former Pioneer co-workers Phil Laing and Tony Liu, of 
The Aerospace Corporation.   Laing and Liu used the independent CHASMP navigation code to look at the Pioneer data.  To within small uncertainties, their result was the same as that obtained by JPL's ODP.   

This resulted in the  Pioneer anomaly Collaboration's discovery paper of 1998  \cite{pioprl}.
Thereafter a detailed analysis 
began using the Pioneer 10 data spanning 3 January 1987 to 22 July 1998 (when the craft was 40 AU to 70.5 AU from the Sun) and Pioneer  11 data spanning 5 January 1987 to 1 October 1990 (when Pioneer 11 was 22.4 to 31.7 AU from the Sun).  It was during this period that we presented our report to CPT-01 conference \cite{cpt01}.

The most significant biases and errors (all in units of $10^{-8}$ cm/s$^2$) are 
radio beam reaction force $(1.10\pm 0.11)$; RTG heat reflected off the craft $(-0.55\pm 0.55)$; differential emissivity of the RTGs $(\pm 0.85)$; non-isotropic radiative cooling of the craft $(\pm 0.48)$; gas leakage $(\pm 0.56)$.    
Note that except for the radio beam reaction force, which mainly is a bias, and the gas leak uncertainty,  the other major systematics are due to heat.  

When added to experimental residuals, this yields the final result, that there is an unmodeled acceleration, approximately towards the Sun, of 
\be
a_P = (8.74 \pm 1.33) \times 10^{-8} \mathrm{cm/s}^2
\ee
which we reported in 2002 \cite{pioprd}.
The conclusion, then, is that this ``Pioneer anomaly" is in the data.  
The question is, ``What is its origin?"\footnote{
One can note a small apparent annual oscillation on top of the constancy in Fig. \ref{correlate}.  Careful analysis of clean, late-time data showed \cite{pioprd} that this signal is indeed significant, {\it as well as} a diurnal signal.  These are believed to be independent of the main anomaly. 

There is another anomaly that is associated with Earth flybys.   
A spacecraft's total {\it geocentric} orbital
energy per unit mass {\it should} be the same before and after the flyby.  
But the data indicates this is not always true \cite{hale}.  
}


\section{Proposed Origins of the Anomaly}
\label{meaning}


{\bf On board systematics:}
It is tempting to assume that radiant heat must be the cause of the acceleration, since only 63 W of directed power could cause the effect (and much more heat than that is available).  

The heat on the craft ultimately comes from the  Radioisotope Thermoelectric Generators (RTGs), which yield heat from the radioactive decay of $^{238}$Pu.  Before launch,
the four RTGs had a total thermal fuel inventory of 2580 W ($\approx
2070$ W in 2002). 
Of this heat 165 W was converted at launch into electrical power emanating from around the main bus ($\approx 65$ W in 2002), significantly from the louvers at the bottom of the bus.

Therefore, it has been argued  whether the anomalous
acceleration is due to anisotropic heat reflection off of the back of the
spacecraft high-gain antennas,  
to the heat emanating from the louvers (open or closed), or if a combination of both these sources must be considered . 

In fact, the craft was designed so that the heat was radiated out in a very fore/aft symmetric manner.   Further, the heat from electric power went down by almost a factor of 3 during the mission.  
With all these points in mind, no one as yet has been able to firmly tie this down, despite heated controversy \cite{piompla}. 
Therefore, heat as a mechanism remains to be clearly resolved, but studies are underway (see Section \ref{searching}).


{\bf Other physics:}
Drag from normal matter dust as well as gravity from the Kuiper belt have been ruled out.  Also, if this is a modification of gravity, it is not universal; i.e., it is not a scale independent force that affects planetary bodies in bound orbits \cite{pioprd}. 
The anomaly could, in principle be i) some modification of gravity, ii) drag from dark matter  or a modification of inertia, or iii) a light acceleration.  (Remember, the signal is a Doppler shift which is only interpreted as an acceleration.)  Future understanding of the anomaly will determine which, if any, of these proposals are viable.

In the above circumstances the true direction of the anomaly should be i) towards the Sun, ii) along the craft velocity vector,
or iii) towards the Earth.  If the origin is heat, or any other spacecraft-generated force, the acceleration would be iv) along the spin axis. (Any internal systematic forces normal to the spin axis are canceled out by the rotation.)


\section{Possibilities for Progress}
\label{searching}


{\bf Studying the entire data set:}  
The major analysis \cite{pioprd} used data from 1987.0 to 1998.5.  However, a long data span would help discern whether the anomaly is truly constant or if there is a time-dependence.  Further, one might be able to do 3-dimensional tracking precisely enough to determine the exact direction of the anomaly.  Perhaps most intriguingly, by closely studying the data around the planetary flybys, it might be determined if there was an onset of the anomaly near these encounters.  

If all the Doppler data, from launch to last contact, could be precisely analyzed together {\it and} all systematics external to the craft could be separated, the above tasks could be accommodated.  The data itself was stored using obsolete formats on obsolete platforms.  Further, it was not to be found in one place, although the NSSDC did have a sizable segment.  The untangling and interpretation of the archival Doppler data is an ongoing and important project. 
Further, there is the telemetry which contains the engineering data, including such things as temperatures, voltages, spin rates, etc.   This data has also now been transferred to modern format.  

In the long run the study of the telemetry might prove to be of most ``use." The Collaboration has long observed that, even if the anomaly turns out to be due to systematics, a thorough anomaly inquiry would still result in a win.  One  would obtain a better understanding of how spacecraft behave in deep space and therefore how to build, model, and track craft there.  

Ultimately things will be decided by the analysis of the Doppler data, id it can show both the size of the anomaly with time and also, with luck, its direction.  This latter will be most difficult to discern because of the large systematics close in to the Sun.    
But the results could be very rewarding.   


{\bf The New Horizons mission to Pluto:}
On 19 Jan 2006 the New Horizons mission to Pluto and the Kuiper Belt was launched from Cape Canaveral.  Although of relatively low mass ($\sim$478 kg), this craft was not designed for precision tracking but it might yield useful information.

The main reason is that for much of its life New Horizons will be in spin-stabilization mode, for example  for much of the period after June 2007 (Jupiter encounter was on 28 Feb. 2007) until soon before the Pluto encounter on  14 July 2015.  This is designed to save fuel so that as much fuel as possible will be available after Pluto encounter to aim at a Kuiper Belt Object.  The Doppler and range data from these periods could supply a test, at some level, of the Pioneer anomaly.    With luck something could be learned from the New Horizons data by 2010 or soon thereafter.    

In summary, we await to see if this mission can be of use to the study of the anomaly.  The analysis of data from the Pluto mission will be a challenge, but though it might not aid our goals, we encourage it.


{\bf A Dedicated Mission?}
If the above efforts are not able to yield a {\it negative} resolution of the anomaly, then a new experimental test might be needed, either as an attached experiment or probe, or even as a dedicated mission.  
In Europe a collaboration arose.  
The aim was to present a proposal for ESA's Cosmic Vision program, with launches during the period 2015-2025.  Its timing would be perfect if the two investigations described above indicate a dedicated test of the anomaly is called for.  

The collaboration decided to submit a two-stage proposal.  The first stage  would use modern 3-axis, drag-free accelerometers  developed by ONERA in France.  This mission would be a slower mission using flybys for gravity assists out to $\sim 10$ AU.  
This mission would get its main power from solar cells.  As such this mission could be flown cheaper and quicker.  But it would have power limitations and entail large solar radiation pressure effects from the solar-cell assemblies that are needed at larger heliocentric distances..  
A later mission would be equipped with modern quantum-technology instruments to measure precise accelerations at longer distances.  
If called for, a dedicated mission could be very exciting and definitive.  


\section{Conclusion}
\label{conclusion} 

Although it is unlikely that the Pioneer anomaly is caused by new physics, that is not ruled out.  It is possible that the anomaly could be something importantly new.  For that reason, and also because we want to be able to account for every source of systematic error in navigation, we continue to pursue this study..  This anomaly will be resolved eventually.

{\it The work described in this manuscript was supported by the US Department of Energy (MMN) and by the National Aeronautics and Space Administration (JDA).}



\end{document}